\documentclass[12pt]{article}
\usepackage{a4wide}
\usepackage{epsfig}
\usepackage{color}

\newcommand{\slk}{/\kern-6pt k}
\newcommand{\sll}{/\kern-4pt l}
\newcommand{\slp}{p\kern-5pt/}
\newcommand{\slq}{q\kern-5.5pt/}
\newcommand{\sls}{s\kern-5.5pt/}

\newcommand{\bea}{\begin{eqnarray}}
\newcommand{\ena}{\end{eqnarray}}
\newcommand{\nn}{\nonumber\\}
\newcommand{\be}{\begin{equation}}
\newcommand{\en}{\end{equation}}
\newcommand{\oone}{\hbox{$1\kern-2.5pt\hbox{\rm l}$}}
\newcommand{\ssigma}{\hbox{$\kern2.5pt\vrule height4pt\kern-2.5pt\sigma$}}

\newcommand{\slell}{/\kern-5pt\ell}

\begin{document}

%file: dubna17/prockoer17\vspace{-12pt}
\begin{flushright}
MITP/17-072\\
%xxxx.xxxxX [hep-ph]\\
%version of \today
\end{flushright}
\vspace{0.5cm}

\begin{center}
{\Large\bf Bread and butter physics: NLO electroweak\\[3pt]
  corrections to polarized top quark decays}\\[0.5cm]
S.~Groote$^1$ and J.G.~K\"orner$^2$\\[1cm]
$^1$ Loodus- ja Tehnoloogiateaduskond, F\"u\"usika Instituut,\\
Tartu \"Ulikool, T\"ahe 4, 51010 Tartu, Estonia\\[7pt]
  $^2$ PRISMA Cluster of Excellence, Institut f\"ur Physik,\\
  Johannes-Gutenberg-Universit\"at, D-55099 Mainz, Germany
\end{center}

\begin{abstract}\noindent
We report on the status of an ongoing calculation of the NLO electroweak
corrections to sequential polarized top quark decays
$t(\uparrow) \to X_b +W^+ (\to \ell^+ +\nu_\ell)$.
\end{abstract}

%%%%%%%%%%%%%%%%%%%%%%%%%%%%%%%%%%%%%%%%%%%%%%%%%%%%%%%%%%%%%%%%%%%%%%%%%%%%
\section{Singly produced top quarks:\\
  Rates, polarization and integrated luminosities}
%%%%%%%%%%%%%%%%%%%%%%%%%%%%%%%%%%%%%%%%%%%%%%%%%%%%%%%%%%%%%%%%%%%%%%%%%%%
The $t$-channel production of single top quarks is mediated by parity-violating
weak interactions which is a necessary condition for a non-vanishing
polarization of the produced top quarks. And, in fact, the polarization of
singly produced top quarks is calculated to be $P_t\sim\,90\,\%$ where the
polarization is primarily along the spectator quark of the production process.
The Standard Model (SM) rates for single top production are given by
\be
\sigma^{8\,{\rm TeV}}=55\cdot 10^3 \,\,fb  \qquad
\sigma^{13\,{\rm TeV}}=136\cdot 10^3 \,fb
\en
Typical numbers on integrated luminosities quoted by the CMS and ATLAS
Collaborations in their analysis of singly produced top quarks are
20 $fb^-1$ @ 8 ${\rm TeV}$ and 3 $fb^-1$ @ 13 ${\rm TeV}$ which
corresponds to $1.1\cdot 10^6$ and $0.41\cdot 10^6$ single top quark events,
respectively. The present experimental situation is nicely reviewed in
Ref.~\cite{Faltermann:2017vry}. The experimental analysis of top quark
polarization by the ATLAS Coll. is based on $20.2\, fb^{-1}\, @\, 8\, {\rm TeV}$
($\sim 1.1 \cdot 10^6$ events)~\cite{Aaboud:2017aqp}. ATLAS quotes a value of
$P_t=0.97\pm0.12$ for the polarization of a singly produced top
quark~\cite{Aaboud:2017aqp}.
%%%%%%%%%%%%%%%%%%%%%%%%%%%%%%%%%%%%%%%%%%%%%%%%%%%%%%%%%%%%%%%%%%%%%%%%%%%%
\section{The hunt is on}
%%%%%%%%%%%%%%%%%%%%%%%%%%%%%%%%%%%%%%%%%%%%%%%%%%%%%%%%%%%%%%%%%%%%%%%%%%%%
The $t\to b$ transition matrix element of an effective current
$J_\mu^{\rm eff} (t \to b)$ is given by (see e.g.\
Ref.~\cite{Bernreuther:2008us})
\be
M_\mu^{\rm eff}=-\frac{g_w}{\sqrt{2}}\bar u_b
\Big\{\gamma_\mu(f_LP_L+f_RP_R)+
\frac{i\sigma^{\mu\nu}q_\nu}{m_W}(g_L P_L+g_R P_R)\Big\}u_t
\en
In the SM one has $f_R,\,g_L, g_R=0$ and $f_L=V^\ast_{tb}\sim 1$. Imaginary
parts of the coupling factors can be generated by final state interactions
(SM; $CP$-conserving) or by introducing non-SM $CP$-violating imaginary
couplings by hand. Real contributions of $f_L,\,f_R,\,g_L, g_R$ to the spin
density elements of the $W^+$ compete with higher order perturbative
corrections. 

Large samples of polarized top quarks are presently being produced at the LHC.
There is a strong ongoing experimental program (ATLAS and CMS) to unravel
the structure underlying polarized top quark decay. In particular, the hunt
is on for signals of non-SM physics in polarized top quark decays.

%%%%%%%%%%%%%%%%%%%%%%%%%%%%%%%%%%%%%%%%%%%%%%%%%%%%%%%%%%%%%%%%%%%%%%%%%%%%
\section{Scope of the problem}
%%%%%%%%%%%%%%%%%%%%%%%%%%%%%%%%%%%%%%%%%%%%%%%%%%%%%%%%%%%%%%%%%%%%%%%%%%%

There are 8 $T$-even and 2 $T$-odd helicity structure function describing
sequential polarized top quark decays
$t(\uparrow) \to X_b+W^+(\to \ell^+ + \nu_\ell)$. Three of these describe
unpolarized top quark decay. As concerns NLO electroweak contributions the
following calculations exist in the literature:
\begin{itemize}
\item[i)]Total unpolarized rate~\cite{Denner:1990ns,Eilam:1991iz}
\item[ii)]Three unpolarized helicity fractions~\cite{Do:2002ky} preceded
  by a soft photon calculation~\cite{Kuruma:1992if}
\item[iii)]Electroweak one-loop corrections to the real and imaginary parts of
  the left and right tensorial coupling factors
  $g_{R,L}$~\cite{GonzalezSprinberg:2011kx,Gonzalez-Sprinberg:2015dea,%
  Arhrib:2016vts,Fischer2017}.
\end{itemize}
The NLO electroweak radiative corrections to 5 $T$-even and to 2 $T$-odd
polarized helicity structure functions are still missing and need to be
calculated.

The electroweak radiative corrections to the unpolarized structure functions
amount to $1.55\%$ in the rate and to $1.3\% - 2.1\%$ for the longitudinal and
transverse structure functions. The NLO electroweak radiative corrections
to the polarized structure functions are not likely to be much larger then
those for the unpolarized case. One will therefore
not win big merits and recognition for such a calculation. We have nevertheless
decided to embark on such a ``bread and butter physics'' enterprise.

%%%%%%%%%%%%%%%%%%%%%%%%%%%%%%%%%%%%%%%%%%%%%%%%%%%%%%%%%%%%%%%%%%%%%%%%%%%%%
\section{Size of radiative corrections to\\ unpolarized structure functions}
%%%%%%%%%%%%%%%%%%%%%%%%%%%%%%%%%%%%%%%%%%%%%%%%%%%%%%%%%%%%%%%%%%%%%%%%%%%%%%
Let us list the radiative corrections to the longitudinal rate $\Gamma_L$
which we normalize to the Born term rate. One has
\be
\hat \Gamma_L=0.703\,\,\Big(1\,-\,
  \underbrace{9.51 \%}_{\rm NLO\ QCD\ \cite{Fischer:1998gsa,Fischer:2001gp}}
  \,-\,\underbrace{3.21 \%}_{\rm NNLO\ QCD\ \cite{Czarnecki:2010gb}}
  \,+\,\underbrace{1.32\%}_{\rm NLO\ EW\ \cite{Do:2002ky}}\Big)
\en
Contrast this with what is known/unknown for the corresponding polarized
longitudinal rate given by
\be
\label{pollong}
\hat \Gamma^P_L=0.703\,\,\Big(1\,-\,
  \underbrace{9.62 \%}_{\rm NLO\ QCD\ \cite{Fischer:1998gsa,Fischer:2001gp}}
  \,-\,\underbrace{?\,\, \%}_{\rm NNLO\ QCD}
  \,+\,\underbrace{?\,\, \%}_{\rm NLO\ EW}\Big)
\en
   
J.G.K. had applied for a grant to the Deutsche Forschungsgemeinschaft (DFG)
proposing, among others, to calculate the NNLO QCD and NLO electroweak
radiative corrections to polarized top quark decays to fill in the missing
pieces such as in Eq.~(\ref{pollong}). The money asked for was a mere pittance.
One of the referees shot down the grant
application with the following argument: ``The polarization of singly produced
top quarks has been measured by the ATLAS Collaboration
with an error of $12\,\%$~\cite{Aaboud:2017aqp}. Why calculate the above radiative corrections which
are likely to be smaller than $12\,\%$?'' My answer: The error on $P_t$ is
based on pre-2016 data. In the meantime the data base has grown by a factor of
7 and will continue to grow in the next few years. The HL-LHC to come in the
next years is expected to collect 250 $fb^-1$ per year and detector which
corresponds to $34\,\times 10^6$ polarized top quarks. The error on the top
quark's polarization is likely to be reduced considerably during the next few
years which makes the calculation of radiative corrections to polarized
top quark decays indispensible.
   
%%%%%%%%%%%%%%%%%%%%%%%%%%%%%%%%%%%%%%%%%%%%%%%%%%%%%%%%%%%%%%%%%%%%%%%%%%%%%%
\section{Counting the number of structure functions}
%%%%%%%%%%%%%%%%%%%%%%%%%%%%%%%%%%%%%%%%%%%%%%%%%%%%%%%%%%%%%%%%%%%%%%%%%%%%%%
The spin density matrix elements of the $W^+$ (also called helicity structure functions) are hermitian. They satisfy
\be
\bigg(H_{\lambda_W\, \lambda'_W}^{\,\lambda^{\phantom x}_t\,\lambda'_t}\bigg)^\dagger=
\bigg(H_{\lambda'_W\, \lambda_W}^{\,\lambda'_t\,\lambda_t}\bigg)
\en
One has to observe the angular momentum constraint $\lambda_W- \lambda'_W=\lambda_t-\lambda'_t$ which
implies
$|\lambda_W- \lambda'_W| \le 1 $.
With the above constraint one counts ten independent double spin density
matrix elements:
\be
H_{++}^{++},\,H_{++}^{--},\,H_{--}^{++},\,H_{--}^{--},\,H_{00}^{++},\,%
H_{00}^{--},\,H_{+0}^{+-},\,H_{0+}^{-+},\,H_{-0}^{-+},\,H_{0-}^{+-}.
\en
We mention that the counting by covariants is not straightforward
because of a nontrivial Schouten identity among the covariants. We sort the
$H_{\lambda_W\, \lambda'_W}^{\,\lambda^{\phantom x}_t\,\lambda'_t}$
into a set of three unpolarized structure functions and a set of seven
polarized structure functions by writing
\bea
H_{++}&=&H_{++}^{++} +H_{++}^{--} \qquad H_{00}=H_{00}^{++} +H_{00}^{--}
\qquad H_{--}=H_{--}^{++} +H_{--}^{--} \nn
H^P_{++}&=&H_{++}^{++} -H_{++}^{--} \qquad H^P_{00}=H_{00}^{++} -H_{00}^{--}
\qquad H^P_{--}=H_{--}^{++} -H_{--}^{--} \nn
H^P_{+0}&=&H_{+0}^{+-} \quad \,H^P_{0+}=H_{0+}^{-+} \quad \,H^P_{-0}=H_{-0}^{-+}
\quad \,H^P_{0-}=H_{0-}^{+-}
\ena
It is convenient to consider particular linear superpositions of the spin
density elements which feature in the angular decay distribution of the
process. These are
\bea
H_U&=&H_{++}  +H_{--} \qquad H_L=H_{00} \qquad H_F=H_{++} -H_{--} \nn 
H^P_U&=&H^P_{++} +H^P_{--} \qquad H^P_L=H^P_{00} \qquad
H^P_F=H^P_{++} -H^P_{--} \nn 
H^P_I&=&\frac{1}{2}\Big({\cal R}e H^P_{+0} -{\cal R}e H^P_{-0}\Big)
\qquad \qquad
H^P_A=\frac{1}{2}\Big({\cal R}e H^P_{+0} +{\cal R}e H^P_{-0}\Big)\nn
H^P_{{\cal I}I}&=&\frac{1}{2}\Big({\cal I}\!m H^P_{+0} -{\cal I}m H^P_{-0}\Big)
\qquad \qquad
H^P_{{\cal I}A}=\frac{1}{2}\Big({\cal I}m H^P_{+0} +{\cal I}m H^P_{-0}\Big)
\ena
%%%%%%%%%%%%%%%%%%%%%%%%%%%%%%%%%%%%%%%%%%%%%%%%%%%%%%%%%%%%%%%%%%%%%%%%%%%%%%
\section{Angular decay distribution}
\begin{figure}\begin{center}
\epsfig{figure=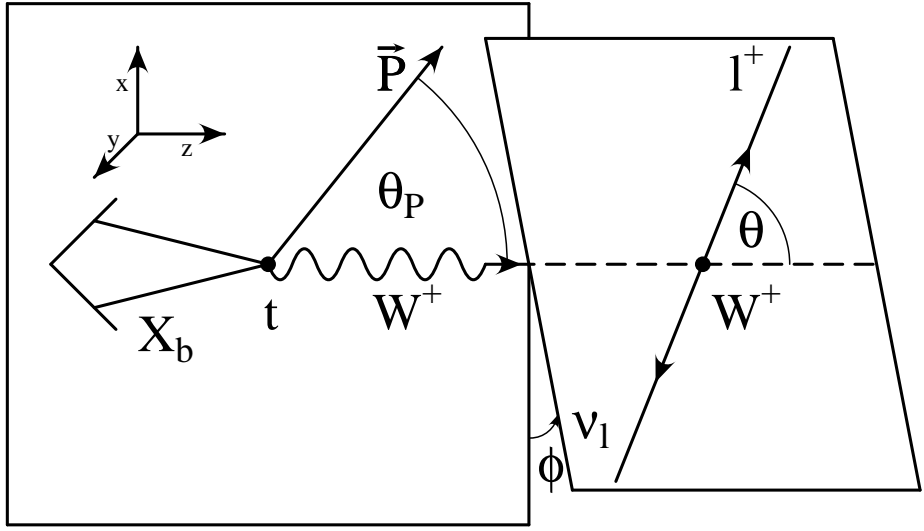, scale=0.7} 
\end{center}
\caption{\label{angdef} Definition of the polar angles $\theta$ and $\theta_P$
  and the azimuthal angle $\phi$ in the sequential decay
  $t(\uparrow) \to X_b +W^+ (\to \ell^+ +\nu_\ell)$.}
\end{figure}

%%%%%%%%%%%%%%%%%%%%%%%%%%%%%%%%%%%%%%%%%%%%%%%%%%%%%%%%%%%%%%%%%%%%%%%%%%%%%
We introduce the polar angles $\theta_P$ and $\theta$ which are defined in the
top quark and $W^+$ rest frames, respectively, as depicted in
Fig.~\ref{angdef}. The azimuthal angle $\phi$ measures the azimuthal angle
between the two planes as shown in Fig.~\ref{angdef}.

The angular decay distribution in the sequential decay
$t(\uparrow) \to X_b +W^+ (\to \ell^+ +\nu_\ell)$ can then be obtained from
the master formula~\cite{Fischer:2001gp}
\begin{equation}
   W(\theta_P,\theta,\phi) \propto
   \sum\limits_{\lambda_W - \lambda^{\prime}_W = \lambda_t - \lambda^{\prime}_t}
   e^{i (\lambda_W - \lambda^{\prime}_W) \phi} \,
   d^1_{\lambda_W 1}(\theta) \, d^1_{\lambda^{\prime}_W 1}(\theta) \,
   H_{\lambda_W \lambda^{\prime}_W}^{\lambda_t \: \lambda^{\prime}_t} \, 
   \rho_{\lambda_t \: \lambda^{\prime}_t} (\theta_P),
 \label{masterformula}
\end{equation}
where the spin density matrix of the top quark is given by
\begin{equation}
   \rho_{\lambda_t \: \lambda^{\prime}_t} (\theta_P) =
   \frac{1}{2} \pmatrix{
   1 + P_t \cos \theta_P & P_t \sin \theta_P \cr
   P_t \sin \theta_P & 1 - P_t \cos \theta_P }.
 \end{equation}
\noindent $ P_t $ is the magnitude of the polarization of the top quark. The
second lower index in the small Wigner $ d(\theta) $-function
$ d^1_{\lambda_W 1} $ is fixed at $ m = 1 $ for zero mass leptons because the
total $ m $-quantum number of the lepton pair along the $ l^{+} $ direction is
$ m = 1 $. One then obtains the normalized angular decay
distribution~\cite{Fischer:2001gp,Bialas:1992ny,Boudreau:2013yna,%
Aguilar-Saavedra:2015yza,Aguilar-Saavedra:2017wpl}
\begin{eqnarray} % *** decay distribution *** %
  \label{angdist}
  && W(\cos \theta,\,\cos \theta_P,\,\phi)  = \frac{1}{4\pi}\,\frac{3}{8}
 \, \bigg\{\Big( \hat H_U + \hat H_U^P \,P_t \cos \theta_P \,\Big) 
   (1 + \cos^2 \theta) \nn && \qquad
   + \Big( \hat H_L +  \hat H_L^P \,P_t \cos \theta_P\Big)2\sin^2 \theta
   + \Big( \hat H_F + \hat H_F^P P_t \cos \theta_P \Big) 2 \cos \theta \nn   
&& \qquad + \hat H_I^P  P_t \sin \theta_P  \,
         2\,\sqrt{2} \sin 2\theta \cos \phi 
          + \hat H_A^P P_t \sin \theta_P \,  
         4\,\sqrt{2} \sin \theta \cos \phi \nn
&& \qquad + \hat H_{{\cal I}I}^P\,P_t \sin \theta_P \,  
         2\,\sqrt{2} \sin 2 \theta \sin \phi 
          + \hat H_{{\cal I}A}^P P_t \sin \theta_P \, 
         4\,\sqrt{2} \sin \theta \sin \phi \nonumber\, \bigg\}
 \end{eqnarray}
where we define hatted normalized structure functions by
\[\hat H^{(P)}_i=\frac{ H^{(P)}_i}{ H_{\rm total}}=H^{(P)}_i/(H_U+H_L)\]
such that $\int d\!\cos\theta\, d\!\cos\theta_P \,d\phi
\,W(\cos \theta,\,\cos \theta_P,\,\phi) =1$.

%%%%%%%%%%%%%%%%%%%%%%%%%%%%%%%%%%%%%%%%%%%%%%%%%%%%%%%%%%%%%%%%%%%%%%%%%%%%%%%
\section{$T$-odd correlations}
%%%%%%%%%%%%%%%%%%%%%%%%%%%%%%%%%%%%%%%%%%%%%%%%%%%%%%%%%%%%%%%%%%%%%%%%%%%%%%%
The last two terms in the angular decay distribution correspond to
$T$--odd angular correlations. This becomes evident by rewriting the
angular factors in terms of the normalized three-vectors of the process
(see Fig.~1):
\be
\hat  P_t= (\sin\theta_P,0,\cos\theta_P)\qquad
\hat  p_\ell=\left( \sin\theta \cos\phi,
             \sin\theta \sin\phi, \cos\theta\right) \qquad
    \hat  q=(0,0,1)
    \en
    One finds
\bea
\sin\theta_P\,\sin\theta\,\sin\phi &=&
\hat q \cdot(\hat P_t \times \hat p_{\ell})
 \\
\sin\theta_P\,\sin2\theta\,\sin\phi &=& 2\,(\hat p_{\ell}\cdot \,\hat q)\,\,
\hat q \cdot(\hat P_t \times \hat p_{\ell}) \nonumber
\ena
Under time reversal $t \to -t$ one has
$(\hat p,\hat P_t) \to (-\hat p,-\hat P_t)$. One therefore calls the 
above two angular correlations $T$--odd correlations. As mentioned before
there are two possible sources of $T$--odd correlations:
\begin{itemize}
\item[i)]SM source: Imaginary parts from absorptive contributions,
\item[ii)]Non-SM source: $CP$-violating imaginary couplings.
\end{itemize}

 %%%%%%%%%%%%%%%%%%%%%%%%%%%%%%%%%%%%%%%%%%%%%%%%%%%%%%%%%%%%%%%%%%%%%%%%%%%%%%
\section{NLO tree-level Feynman diagrams}
 %%%%%%%%%%%%%%%%%%%%%%%%%%%%%%%%%%%%%%%%%%%%%%%%%%%%%%%%%%%%%%%%%%%%%%%%%%%%%
There are four NLO electroweak tree level Feynman diagrams that contribute to 
$t \to b + W^+ + \gamma$ since the photon can couple to the top quark, the
bottom quark and the $W^+$-boson (three diagrams) plus the
diagram where the internal $W^+$ is replaced by the Goldstone boson $\xi^+$.
The procedure to calculate the tree-level diagrams is standard. One splits the
tree-level matrix element into a hard and a soft photon part according to
\be
M^\mu M^{\nu \ast}({\rm hard+soft})=\underbrace{\Big(M^\mu M^{\nu \ast}
({\rm hard+soft})-|M|^2({\rm soft})\Big)}_{\rm IR \,and\, M \,safe}
+\underbrace{M^\mu M^{\nu \ast}({\rm soft})}_{\rm universal}
\en
The first piece is infrared and collinear finite and can be safely integrated
in $D=4$ dimension. The second piece is universal in the sense that it is
proportional to the Born term. The finite piece has to be projected onto the
relevant eight $T$-even structure functions and then has to be integrated over
the photon phase space. This has been done. For the projection onto the
structure functions it is convenient to peruse the eight covariant projectors
listed in Ref.~\cite{Fischer:2001gp}. The second universal piece has been
known since long.
%%%%%%%%%%%%%%%%%%%%%%%%%%%%%%%%%%%%%%%%%%%%%%%%%%%%%%%%%%%%%%%%%%%%%%%%%%%%%
\section{NLO electroweak one-loop vertex graphs}
%%%%%%%%%%%%%%%%%%%%%%%%%%%%%%%%%%%%%%%%%%%%%%%%%%%%%%%%%%%%%%%%%%%%%%%%%%%%%
In the Feynman 't Hooft gauge there are altogether 18 NLO electroweak
one-loop Feynman vertex diagrams that
contribute to $t \to b + W^+ $. The corresponding one-loop integrals are not
simple in that they have
five mass scales: {$m_t,\,m_b,\,m_W,\,m_Z,\,m_H$}.

In the following we shall set $m_b=0$. The number of vertex diagrams is
reduced to 13 since $g_{Hbb}=0$ and $g_{\chi_3bb}=0$. The complexity of the
one-loop integrals is also somewhat reduced because now there are only four mass
scales. The calculation of the 13 one-loop diagrams plus the counter-term
diagrams needed for renormalization has been done.

In the limit $m_b=0$ one has the simplifying feature that the longitudinal
and transverse projection of the matrix element $M_\mu$ are proportional
to the corresponding projections of the Born matrix element, i.e. one has
($x=m_W/m_t$)~\cite{Kuruma:1992if}
\bea
\label{lproject}
    \varepsilon^{\ast\,\mu}(0) M_\mu&=&\Big(f_L -x\, g_R\Big)
    \,\, \varepsilon^{\ast\,\mu} (0) M_\mu(Born) \nn
        \varepsilon^{\ast\,\mu}(-) M_\mu&=&\Big(f_L-\frac{1}{x}\,\,g_R\Big)\,\, \varepsilon^{\ast\,\mu} (-) M_\mu(Born)
   \ena
%%%%%%%%%%%%%%%%%%%%%%%%%%%%%%%%%%%%%%%%%%%%%%%%%%%%%%%%%%%%%%%%%%%%%%%%%%%%%%%
\section{Electroweak one-loop vertex graphs\\ that admit of absorptive cuts}
%%%%%%%%%%%%%%%%%%%%%%%%%%%%%%%%%%%%%%%%%%%%%%%%%%%%%%%%%%%%%%%%%%%%%%%%%%%%%%%
%\vspace{0.5cm}
\begin{figure}
\begin{center}
  \epsfig{file=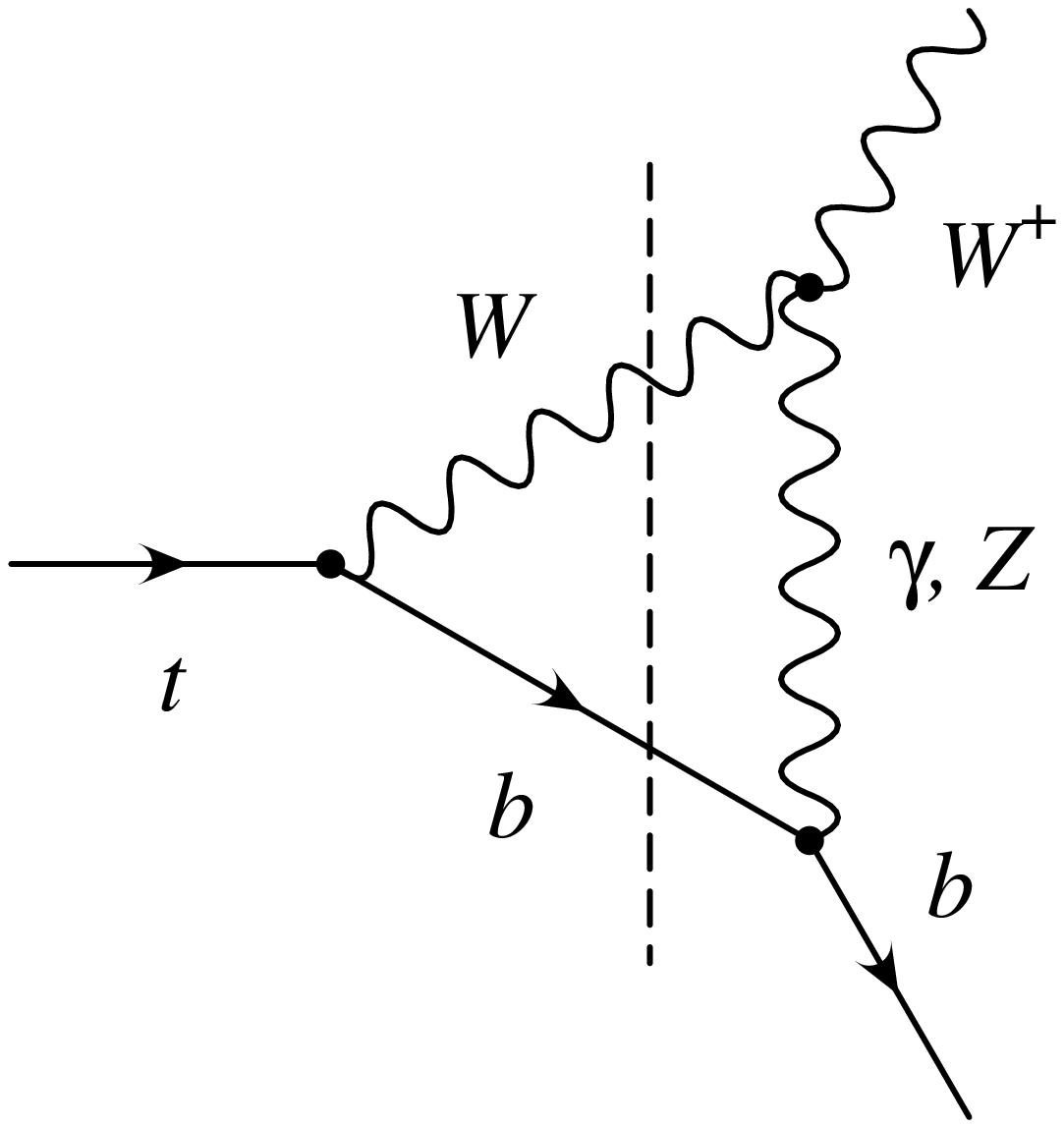, scale=0.5} \qquad
  \epsfig{file=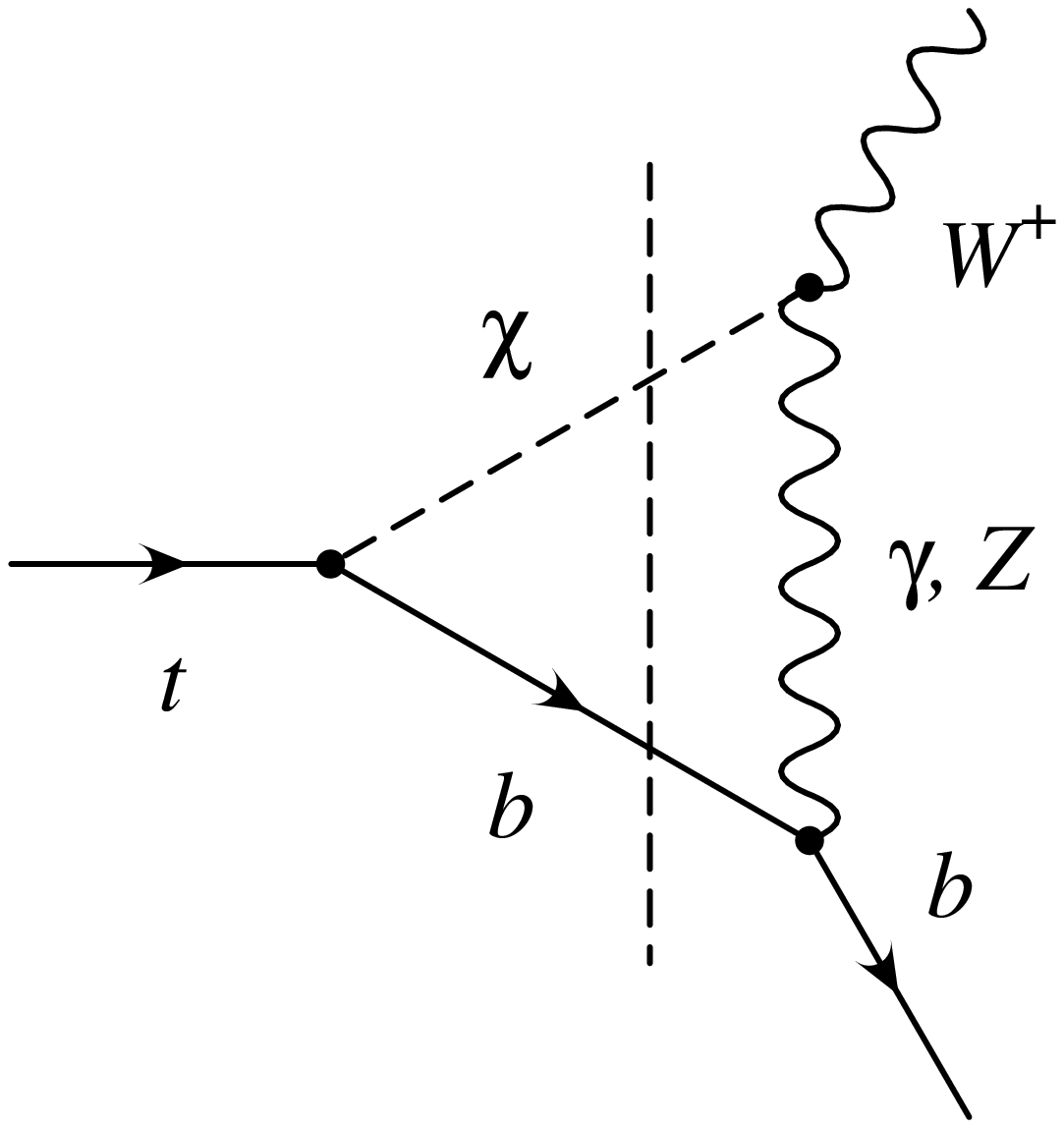, scale=0.5} 
\end{center}
\caption{\label{figloop3}Absorptive parts of the four Feynman diagrams that
  contribute to the $T$-odd correlations in polarized top decays.}
\end{figure}
There are four $(m_B=0)$ NLO electroweak one-loop Feynman diagrams for
$t \to b + W^+ $ that admit of absorptive cuts (often also referred to as
final state interactions/rescattering corrections) which we show in
Fig.~\ref{figloop3}. We have carefully extracted the imaginary parts of the
four diagram in analytical form. Numerically one obtains
\be
{\cal I}m\,g_R=-0.217\,\%
\en

The two $T$-odd contributions can be projected out by taking the moment of the
decay distribution w.r.t.\ $\sin\phi$. The $T$-odd contribution
$\hat H_{{\cal I}A}^P$ can be isolated by limiting the $\phi$-integration to
the interval $[0,\,\pi]$. One obtains
\be
\int_{-1}^1d\!\cos\theta\int_{-1}^1d\!\cos\theta_P\int_0^\pi d\phi\,
  W(\cos \theta,\,\cos \theta_P,\,\phi)=\frac{1}{2}
  \left(1+\frac{3\pi}{4\sqrt{2}}\hat H_{{\cal I}A}^P P_t\right)
\en
where we now use the Born term total structure function $B_{U+L}$ as
normalizing denominator. For the ratio of structure functions one obtains
\be
\hat H_{{\cal I}A}^P=\frac{H^P_{{\cal I}A}}{B_{U+L}}
  =-\frac{1}{\sqrt{2}}\frac{(1-x^2)}{(1+2x^2)}\,
  {\rm Im} g_R=-0.1965 \,{\rm Im} g_R = 0.0427 \%
\en
The azimuthal analyzing power is thus given by a tiny
\be
\hat H_{{\cal I}A}^P\,\frac{3\pi}{4\sqrt{2}}=0.0711\%
\en

%%%%%%%%%%%%%%%%%%%%%%%%%%%%%%%%%%%%%%%%%%%%%%%%%%%%%%%%%%%%%%%%%%%%%%%%%%%%%
\section{Final remarks}
%%%%%%%%%%%%%%%%%%%%%%%%%%%%%%%%%%%%%%%%%%%%%%%%%%%%%%%%%%%%%%%%%%%%%%%%%%%%%
Let us discuss some of the features of the NLO electroweak corrections to
sequential polarized top quark decays treated in this talk. The soft photon
contributions are proportional to the Born terms with a universal
proportionality factor $H^{(P)}_i({\rm soft})= a B^{(P)}_i$ while the hard
photon contributions are {\it not} proportional to the Born terms.
In the limit $m_b=0$ we have found several simplifications for the
contributions of the one-loop vertex graphs. First, the number of contributing
NLO one-loop vertex graphs is reduced. Second, the structure functions
$H_{--},\,H^P_{--},\,H^P_{-0}$ and $H^P_{0-}$ vanish, and third, the
contributions of the one-loop vertex graphs to the structure functions are
proportional to the corresponding Born terms with non-universal
proportionality factors $H^{(P)}_i(\hbox{one-loop})=b_i\,B^{(P)}_i$.
The non-universal factors $b_i$ are not difficult to obtain.

The happy news is that we have assembled all ingredients necessary to obtain
the NLO electroweak corrections to sequential polarized top quark decays.
What remains to be done is to put the various pieces together and to evaluate
them numerically. A first result was reported on that the azimuthal
analyzing power of the SM absorptive $T$-odd contribution is a tiny $0.07\,\%$.
If experimentalists discover large $T$-odd effects in polarized top quark
decays they would be most certainly due to non-SM $CP$-violating imaginary
coupling terms with only a tiny contamination from SM absorptive contributions.
%%%%%%%%%%%%%%%%%%%%%%%%%%%%%%%%%%%%%%%%%%%%%%%%%%%%%%%%%%%%%%%%%%%%%%%%%%%%%
\subsection*{Acknowledgments}
%%%%%%%%%%%%%%%%%%%%%%%%%%%%%%%%%%%%%%%%%%%%%%%%%%%%%%%%%%%%%%%%%%%%%%%%%%%%%%
This work was supported by the Estonian Science Foundation under grant
No.~IUT2-27. S.G.\ acknowledges the hospitality and support of the theory
group THEP at the Institute of Physics at the University of Mainz.
J.G.K.\ would like to thank the Heisenberg-Landau Foundation for financial
support and M.A.~Ivanov for logistic support in the broadest sense and his
fine hospitality while J.G.K.\ was visiting Dubna.
%%%%%%%%%%%%%%%%%%%%%%%%%%%%%%%%%%%%%%%%%%%%%%%%%%%%%%%%%%%%%%%%%%%%%%%%%%%%%%
%%%%%%%%%%%%%%%%%%%%%%%%%%%%%%%%%%%%%%%%%%%%%%%%%%%%%%%%%%%%%%%%%%%%%%%%%%%%%%

\end{document}